\documentclass[preprint,3p]{elsarticle}

\usepackage{amssymb}
\usepackage[intlimits,sumlimits]{amsmath}
\usepackage{lineno}
\usepackage{subcaption}
\biboptions{sort&compress}

\journal{Physics Letters B}

\begin{document}
%\linenumbers
\begin{frontmatter}
 \title{Measurement of the space-like $\pi^0$ transition form factor }
 \author{
M.~Ablikim$^{1}$, M.~N.~Achasov$^{4,c}$, P.~Adlarson$^{77}$, X.~C.~Ai$^{82}$, R.~Aliberti$^{36}$, A.~Amoroso$^{76A,76C}$, Q.~An$^{73,59,a}$, Y.~Bai$^{58}$, O.~Bakina$^{37}$, Y.~Ban$^{47,h}$
% ...
\\
\vspace{0.2cm}
(BESIII Collaboration)\\
\vspace{0.2cm} {\it
$^{1}$ Institute of High Energy Physics, Beijing 100049, People's Republic of China\\
% ...
$^{4}$ Budker Institute of Nuclear Physics SB RAS (BINP), Novosibirsk 630090, Russia\\
% ...
$^{36}$ Johannes Gutenberg University of Mainz, Johann-Joachim-Becher-Weg 45, D-55099 Mainz, Germany\\
$^{37}$ Joint Institute for Nuclear Research, 141980 Dubna, Moscow region, Russia\\
% ...
$^{47}$ Peking University, Beijing 100871, People's Republic of China\\
% ...
$^{58}$ Southeast University, Nanjing 211100, People's Republic of China\\
$^{59}$ State Key Laboratory of Particle Detection and Electronics, Beijing 100049, Hefei 230026, People's Republic of China\\
% ...
$^{73}$ University of Science and Technology of China, Hefei 230026, People's Republic of China\\
% ...
$^{76}$ University of Turin and INFN, (A)University of Turin, I-10125, Turin, Italy; (B)University of Eastern Piedmont, I-15121, Alessandria, Italy; (C)INFN, I-10125, Turin, Italy\\
$^{77}$ Uppsala University, Box 516, SE-75120 Uppsala, Sweden\\
% ...
$^{82}$ Zhengzhou University, Zhengzhou 450001, People's Republic of China\\
\vspace{0.2cm}
$^{a}$ Deceased\\
% ...
$^{c}$ Also at the Novosibirsk State University, Novosibirsk, 630090, Russia\\
% ...
$^{h}$ Also at State Key Laboratory of Nuclear Physics and Technology, Peking University, Beijing 100871, People's Republic of China\\
% ...
}
 }
 \begin{abstract}
 Based on $2.93\,\text{fb}^{-1}$ of $e^+e^-$ collision data taken with the BESIII detector at a center-of-mass energy of 3.773\,GeV, the two-photon fusion process $e^+e^-\to e^+e^-\pi^0$ is investigated using a single-tag approach. The differential Born cross section $\text{d}\sigma/\text{d}Q^2$ and the space-like transition form factor $|F(Q^2)|$ of the $\pi^0$ are measured as  functions of the squared momentum transfer $Q^2$ of the tagged, scattered lepton. The measurement covers the range $0.2 < Q^2 < 3.5\,\text{GeV}^2$. The results are consistent with previous measurements, and provide a significant improvement for $Q^2<2\,\text{GeV}^2.$
 \end{abstract}



\end{frontmatter}


\section{Introduction}
The interaction of two photons with electrically neutral hadrons is described by transition form factors (TFFs). These form factors are functions $F(Q_1^2,Q_2^2)$ of the virtualities $Q_i^2 = -q_i^2$ of the two photons and provide information about the internal structure of the hadrons. Recently, interest in TFFs of light mesons has increased due to their relevance to the Standard Model (SM) prediction of the anomalous magnetic moment of the muon $a_\mu= \frac{(g-2)_\mu}{2}$, where the hadronic light-by-light scattering (HLbL) contributes one of the dominant sources of theoretical uncertainty~\cite{Aoyama:2020ynm,Aliberti:2025beg,Melnikov:2003xd,Masjuan:2017tvw,Colangelo:2017fiz,Hoferichter:2018kwz,Gerardin:2019vio,Bijnens:2019ghy,Colangelo:2019uex,Pauk:2014rta,Danilkin:2016hnh,Jegerlehner:2017gek,Knecht:2018sci,Eichmann:2019bqf,Roig:2019reh,Blum:2019ugy,Colangelo:2014qya}. Further improvements in the SM prediction are necessary to match the accuracy of 124\,ppb of the world average of direct measurements of $a_\mu$~\cite{Muong-2:2025xyk}.

The most recent estimate of the HLbL contribution is based on data-driven approaches using dispersion relations, which help reduce the model dependence of previous estimates. The dominating process in HLbL is the exchange of light pseudoscalar mesons, predominantly the $\pi^0$, which is described by the respective TFFs. It was demonstrated in Ref.~\cite{Nyffeler:2016gnb} that the most relevant contributions from the $\pi^0$ TFF for the calculation of $a_\mu^{\rm HLbL}$ come from virtualities of $Q_i^2\leq 1\,\text{GeV}^2$.

\begin{figure}
  \centering
  \begin{subfigure}[b]{0.45\textwidth}
    \includegraphics[width=\textwidth]{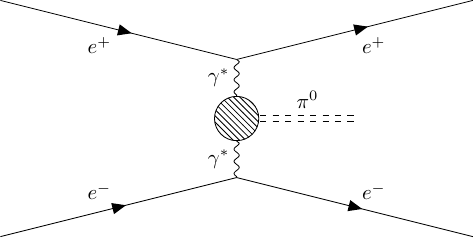}
    \caption{Space-like fusion}
    \label{fig:feynt}
  \end{subfigure}
  \hfill
  \begin{subfigure}[b]{0.45\textwidth}
    \includegraphics[width=\textwidth]{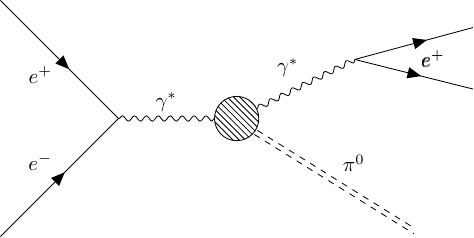}
    \caption{Time-like production}
    \label{fig:feyns}
  \end{subfigure}
    \caption{Feynman diagrams of the space-like two-photon fusion (a) and time-like radiative production (b) amplitudes in $e^+e^-\to e^+e^-\pi^0$. Shaded blobs denote the vertices where the TFF $F(Q_1^2,Q_2^2)$ enters.}
\end{figure}

Available data on the space-like $\pi^0$ TFF have been obtained from measurements of two-photon fusion reactions at $e^+e^-$ colliders, as illustrated in Fig.~\ref{fig:feynt}, where the cross section is directly proportional to the absolute square of the TFF. The cross section shows a strong dependence on the virtualities of the exchanged photons, which are equivalent to the momentum transfers $Q_i^2$ of the scattered electrons\footnote{Unless specified otherwise, the term ``electrons'' refers to both electrons and positrons.}. Thus, measurements are typically performed in a single-tag strategy, in which one photon is required to be quasi-real $(Q_2^2\approx 0\,{\rm GeV}^2)$ and the associated electron is assumed to escape detection at very small scattering angles. The other photon carries a sizable virtuality $Q_1^2 \approx 4E_1E_1^\prime\sin^2(\theta_1/2)$, which is deduced from the scattering angle $\theta_1$, the initial beam energy $E_1$, and the detected energy $E_1^\prime$ of the scattered electron. For simplicity, the quasi-real photon is omitted from the discussion, so that the TFF $F(Q_1^2,Q_2^2\approx0)$ is presented as $F(Q^2)$. Currently, the world data on the $\pi^0$ TFF are dominated by the results obtained at the B-factories BaBar and Belle~\cite{BaBar:2009rrj,Uehara:2012ag}, which primarily cover the large momentum transfer region $3 \leq Q^2 \leq 40\,\text{GeV}^2$. At lower $Q^2$, the results of the CELLO~\cite{Behrend:1990sr} and CLEO~\cite{Gronberg:1997fj} experiments are available covering virtualities down to $0.5\,\text{GeV}^2$, however, with relatively large uncertainties especially at the lowest $Q^2$ values.

The lack of high-precision data has triggered the construction of the $\pi^0$ TFF using dispersive methods by exploiting available time-like data from meson decays and hadronic cross sections~\cite{Hoferichter:2018dmo}. In addition, competitive lattice QCD calculations of meson TFFs are available~\cite{Gerardin:2019vio}. Both approaches contribute to the current consensus on the HLbL contribution to $a_\mu$, but experimental confirmation is still required.

This Letter reports a new measurement of the space-like $\pi^0$ TFF using two-photon fusion events collected with the BESIII detector in $e^+e^-$ collisions. Based on data taken at a center-of-mass (CM) energy of $\sqrt{s}=3.773\,\text{GeV}$, a single-tag analysis is performed to provide precise measurements of the $\pi^0$ TFF in the  momentum transfer region relevant for the HLbL contribution to $a_\mu$.

\section{Detector and Data}
The BESIII detector~\cite{Ablikim:2009aa} records symmetric $e^+e^-$ collisions provided by the BEPCII storage ring~\cite{Yu:2016cof}, which operates with a peak luminosity of $1\times10^{33}$\,cm$^{-2}$s$^{-1}$ in the CM energy range $1.84 \leq \sqrt{s}\leq4.95\,\text{GeV}$. BESIII has collected large data samples in this energy region~\cite{BESIII:2020nme}. The cylindrical core of the BESIII detector covers 93\% of the full solid angle and consists of a helium-based multilayer drift chamber~(MDC), a plastic scintillator time-of-flight system~(TOF), and a CsI(Tl) electromagnetic calorimeter~(EMC), which are all enclosed in a superconducting solenoidal magnet providing a 1.0\,T magnetic field. The solenoid is supported by an octagonal flux-return yoke with resistive plate counter muon identification modules interleaved with steel.
The charged-particle momentum resolution at $1\,{\rm GeV}/c$ is $0.5\%$, and the ${\rm d}E/{\rm d}x$ resolution is $6\%$ for electrons from Bhabha scattering. The EMC measures photon energies with a resolution of $2.5\%$ ($5\%$) at $1$\,GeV in the barrel (end cap) region. The time resolution in the TOF barrel region is 68\,ps, while that in the end cap region is 110\,ps.

The data analyzed in this letter have been acquired in 2011 and 2012 and correspond to an integrated luminosity of $(2931.8\pm0.2_{\rm stat}\pm13.8_{\rm syst})\,\text{pb}^{-1}$~\cite{Ablikim:2015orh}. Simulated data samples are produced using a {\sc geant4}-based~\cite{Agostinelli:2002hh} Monte Carlo (MC) package, which includes the geometric description of the BESIII detector and the detector response. These samples are used to determine detection efficiencies and to estimate background contributions. The simulation accounts for the beam energy spread and initial state radiation (ISR) in the $e^+e^-$ annihilations with the {\sc kkmc} generator~\cite{Jadach:1999vf,Jadach:2000ir}. The inclusive MC sample includes the production of $D\bar{D}$ pairs (including quantum coherence for the neutral $D$ channels), the non-$D\bar{D}$ decays of the $\psi(3770)$, the ISR production of the $J/\psi$ and $\psi(3686)$ states, and the continuum processes incorporated in {\sc kkmc}.

Two-photon collision reactions are simulated using the {\sc Ekhara\,3.1} event generator, an update of Ref.~\cite{Czyz:2018jpp}. The simulation of two-photon production of light pseudoscalar mesons uses exact matrix elements and includes the full radiative corrections at next-to-leading order (NLO). The TFF is modeled using a three-octet framework~\cite{Czyz:2017veo}, while a simple vector meson dominance (VMD) based model is used for consistency checks. {\sc Ekhara\,3.1} also allows for the generation of $\pi^0\pi^0$ final states according to the dispersive analysis of Danilkin, \textit{et al.}~\cite{Danilkin:2018qfn}.

\section{Analysis}
We adopt a single-tag strategy to study the TFF $F_\pi(Q_1^2,Q_2^2=0)$ as a function of a single virtuality $F_\pi(Q^2)$. To establish $\pi^0$ production by two-photon fusion in $e^+e^-$ collisions requires the detection of one of the scattered electrons and at least two photons from the decay of the $\pi^0$.

Charged tracks detected in the MDC are required to be within a polar angle ($\vartheta$) range of $|\cos\vartheta|<0.93$, where $\vartheta$ is defined with respect to the $z$-axis, which is the symmetry axis of the MDC. The distance of closest approach to the interaction point (IP) of each track must be less than 10\,cm along the $z$-axis, $|V_{z}|$, and less than 1\,cm in the transverse plane, $|V_{xy}|$. Tracks are identified as electrons if the ratio of energy deposited in the EMC to the momentum measured in the MDC is larger than $0.8$. Only events with exactly one track that satisfies the above criteria and is identified as an $e^\pm$ candidate are retained for further analysis.

Photon candidates are identified using showers in the EMC.  The deposited energy of each shower must be more than 25\,MeV in the barrel region ($|\cos \vartheta|< 0.80$) and 50\,MeV in the end cap region ($0.86 <|\cos \vartheta|< 0.92$). To exclude showers that originate from charged tracks, the angle between the EMC shower and the nearest charged track at the EMC, measured from the IP, must be greater than 10 degrees. To suppress electronic noise and showers unrelated to the event, the difference between the EMC time and the event start time is required to be within [0, 700]\,ns. Events with fewer than two photon candidates are rejected. For events with more than two, all unique photon pairings are tested in the subsequent signal selection procedure.

The single-tag strategy assumes that the undetected electron is scattered with negligible momentum transfer, corresponding to the exchanging a quasi-real photon. The undetected electron is reconstructed from the measured particles using four-momentum conservation for the reaction $e^+e^-\to e^+e^-\pi^0$. To ensure the small momentum transfer, the inferred scattering angle $\theta$ of the undetected electron is required to satisfy $\cos\theta > 0.99$.

Assuming the selected photon pairs stem from a $\pi^0$ decay, the helicity angle $\theta_H$ distribution is expected to be flat. The angle $\theta_H$ is defined as the angle between one of the photons in the rest frame of the pion candidate and the direction of motion of that pion in the CM frame. A strong peak at large values of $|\cos\theta_H|$ is observed, which is attributed to radiative Bhabha scattering events. To suppress this background, events with $|\cos\theta_H|>0.8$ are rejected.

The parameter $R_\gamma = (\sqrt{s} - p^* - E^*)/\sqrt{s}$, with $p^*=|\vec p^{\,*}_{e^\pm} + \vec p^{\,*}_{\pi^0}|$ and $E^* = E^*_{e^\pm} + E^*_{\pi^0}$ representing the magnitude of momentum and the energy of the reconstructed $e^\pm \pi^0$ system in the CM frame, respectively, is originally derived by energy-momentum conservation to suppress events with initial state radiation~\cite{BaBar:2009rrj}. Here, it is found that requiring $R_\gamma\leq0.15$ is effective in suppressing background contributions from incompletely reconstructed hadronic final states produced in $e^+e^-$ annihilation, continuum contributions, as well as decays of vector charmonia. 

Radiative Bhabha scattering constitutes the dominant background, although it exhibits a non-peaking distribution in $m(\gamma\gamma)$. Peaking background comes from other two photon processes containing at least one $\pi^0$ in the final state. These contributions are suppressed by requiring the sum of energy deposits of additional photons in the events to satisfy $\sum E_\gamma^{\rm extra}\leq 0.17\,\text{GeV}$.

\begin{figure}[htb]
 \centerline{\includegraphics[width=0.8\textwidth]{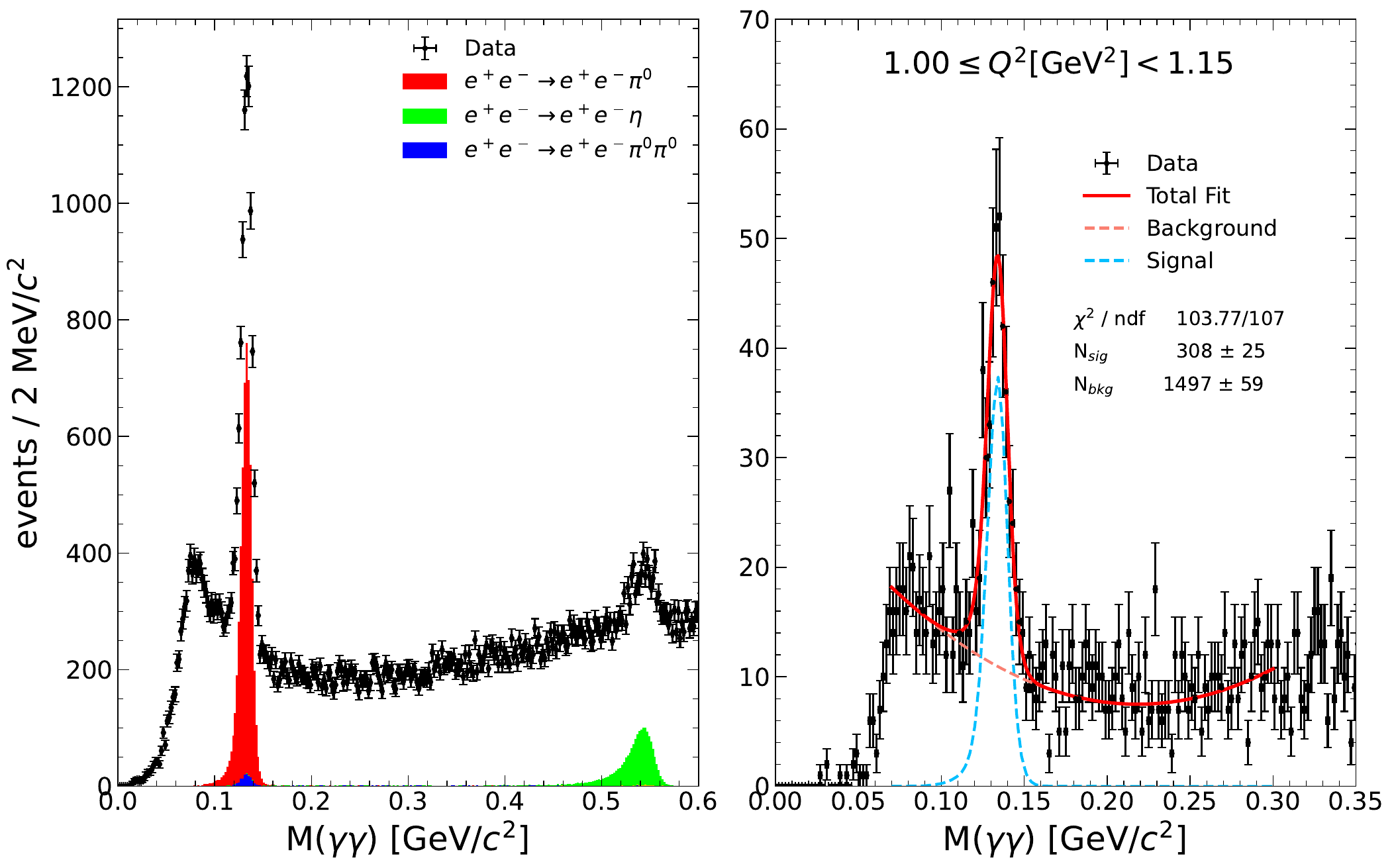}}
 \caption{\label{fig:mass}The distributions of $M(\gamma\gamma)$ after applying all requirements for all events (left) and for events in one $Q^2$ interval (right). Data are shown with error bars, MC distributions of two-photon production channels ($\pi^0$ [red], $\eta,\eta^\prime$ [green], $\pi^0\pi^0$ [blue]) are shown with shaded and stacked histograms. The remaining contributions are attributed to radiative Bhabha scattering events. The red curve shows the fit result of the signal yield determination along with signal [blue] and background components [yellow]. }
\end{figure}

If more than one combination of two photons per event satisfies all the above criteria, the combination with the smallest transverse momentum $p_t^*$ of the measured $e^\pm \pi^0$ system in the CM frame is retained. This affects 0.02\% of the selected events.

In single-tag events, the momentum transfer of the untagged electron is supposed to be small, $Q^2_2\approx0\,\text{GeV}^2$, as anticipated for quasi-real photons. The inferred values of $Q^2_2$ are found to be below $0.28\,\text{GeV}^2$ for 95\% and below $0.49\,\text{GeV}^2$ for 99.7\% of the events within a $3\sigma$ mass window around the $\pi^0$ peak. According to simulation, 99.7\% of the reconstructed events originate from generated values with $Q^2_2\leq0.06\,\text{GeV}^2$, reflecting the limited resolution in the reconstruction of the kinematics of the untagged electron.

Figure~\ref{fig:mass} shows the invariant mass distribution $M(\gamma\gamma)$ of the selected photon pairs after all selection criteria have been applied. A large $\pi^0$ signal is visible above combinatorial background. A clear peak from reconstructed $\eta$ decays is also present. To investigate the $\pi^0$ TFF as a function of $Q^2$, the signal yield is determined by fitting the $\pi^0$ peak in the invariant mass distribution in each momentum transfer interval. The selected events span a range of $0.2\leq Q^2\leq3.5\,\text{GeV}^2$. The interval sizes are chosen to be identical with previous results~\cite{Gronberg:1997fj} at high $Q^2$ to allow for direct comparison.  The interval sizes are typically an order of magnitude larger than the $Q^2$ resolution. Thus, migration effects between $Q^2$ intervals caused by the resolution are negligible.

Peaking background contributions visible in the integrated mass spectrum of Fig.~\ref{fig:mass} are found to be at the sub-percent level in individual $Q^2$ intervals. Though neglected in the yield extraction, their potential effect is accounted for in the systematic uncertainty. Possible irreducible background from the time-like process depicted in Fig.~\ref{fig:feyns} is also neglected, as its cross section is approximately three orders of magnitude lower than that of the space-like signal process.

The fits for each $Q^2$ interval are performed in the mass region $0.07\leq M(\gamma\gamma)\leq0.30\,\text{GeV}/c^2$. The signal shape is modeled using the MC simulated mass distribution convolved with a Gaussian function to account for possible differences in detector resolution between data and simulation. The background is described by a second-order polynomial function. Figure~\ref{fig:mass} illustrates the fit for an individual $Q^2$ interval.

The modulus squared of the TFF $F^2(Q^2)$ is determined from the differential Born cross section for $\pi^0$ production as follows:
 \begin{equation}\label{eq:TFF}
  |F(Q^2)|^2 = \frac{{\rm d}\sigma_{\rm Born}}{{\rm d}Q^2}\left(\frac{{\rm d}\sigma_{\rm WZW}}{{\rm d}Q^2}\right)^{-1} = \frac{N_{\rm sig}}{\Delta Q^2 \, \mathcal{L} \, \varepsilon \, (1+\delta)\, \mathcal{B}_{\gamma\gamma}}\left(\frac{{\rm d}\sigma_{\rm WZW}}{{\rm d}Q^2}\right)^{-1},
 \end{equation}

where $N_{\rm sig}$ is the signal yield, $\Delta Q^2$ is the interval size, $\mathcal{L}$ is the integrated luminosity, $\varepsilon$ is the reconstruction efficiency, $(1+\delta)$ is the radiative correction factor, $\mathcal{B}_{\gamma\gamma}$ is the branching fraction of $\pi^0\to\gamma\gamma$, and $\sigma_{\rm WZW}$ is the point-like Born cross section. The individual values of each term in Eq.~\ref{eq:TFF} and the results are listed in Tab.~\ref{tab:final} and illustrated in Fig.~\ref{fig:final}.

The efficiency $\varepsilon$ is determined using MC simulations as the ratio of reconstructed to generated signal events. Slight variations in $\varepsilon(Q^2)$ are observed depending on the charge of the tagged electron, which correlates with the initial beam directions. The average efficiency in each $Q^2$ interval is applied, and the deviation from the average is included in the systematic uncertainty. Radiative effects are corrected using the ratio of cross section obtained from the signal MC samples including NLO radiative corrections to those from the Born-level process. The point-like Born cross section $\sigma_{\rm WZW}$ is taken from the MC simulation of the signal process using only the Wess-Zumino-Witten~\cite{Wess:1971yu,Witten:1983tw} term and assuming a constant TFF. In the simulation, the second virtuality is restricted to $Q^2\leq0.49\,\text{GeV}^2$, consistent with the reconstructed kinematic range.

When reconstructing the TFF from simulated events and comparing it to the model input under the ideal single-tag assumption $|F(Q_1^2,0)|$, an almost constant offset of approximately 2\% is observed in the region $0.3\leq Q^2\leq3.5\,\text{GeV}^2$ in all tested TFF models. This offset is found to arise from the non-zero second virtuality in the reconstructed events, which causes a slight underestimation of the analytic model. Therefore, a 2\% correction is applied to the TFF measurement.

\section{Systematic Uncertainties}
Several aspects of the data-taking conditions, as well as reconstruction and analysis procedures, are tested for systematic effects on the result presented in this Letter. In the following, the sources found to have significant impact are discussed in detail.

The integrated luminosity is determined using large angle Bhabha scattering with an accuracy of $0.5\%$~\cite{Ablikim:2015orh}. Apart from a negligible fraction, all selected events are triggered by both a single track trigger and a two-cluster energy threshold trigger. The uncertainty of the latter is estimated to be 0.2\% for events with EMC clusters depositing at least 0.2\,GeV~\cite{Berger:2010my}, which is assigned as the uncertainty due to the trigger.

The tracking efficiency for electrons is studied using radiative Bhabha scattering events in Ref.~\cite{BESIII:2018hqu}. An uncertainty of 1\% per track is assigned. Using the same control sample, the uncertainty of PID using the ratio of deposited energy and measured momenta is studied in Ref.~\cite{BESIII:2019ldo}. The average uncertainty weighted by momenta and polar angles is less than 1\% per track.

The photon reconstruction efficiency is studied using radiative muon pair production events as described in Ref.~\cite{Prasad:2016gxm}. An uncertainty of 1\% per photon is assigned.

The relative uncertainty of the $\pi^0$ branching fraction is 0.03\%~\cite{ParticleDataGroup:2024cfk}, and is taken as the corresponding systematic uncertainty.

To evaluate the impact of analysis conditions, relevant observables are randomly varied within a range of $\pm 3\sigma$ corresponding to their respective resolutions. All conditions are varied independently and simultaneously over a set of 1000 samples. The widths of the resulting distributions of efficiency-corrected signal yield in each $Q^2$ interval are taken as the systematic uncertainties. The individual contributions range from $0.6\%$ to $38\%$; the largest values occur for the first and the last two $Q^2$ bins.  

The statistical uncertainties of the reconstruction efficiency as well as the maximum deviations of the $e^-$ and $e^+$ tagged reconstruction efficiencies vs.~$Q^2$ from the average values are considered as the systematic uncertainties. In addition, the model dependence of the reconstruction efficiency and radiative corrections is tested by repeating the analysis using simulations based on a simple VMD model for the TFF. The relative differences in the resulting Born cross sections, which range from $0.6\%$ to $7.4\%$, are taken as the systematic uncertainties.

The uncertainty of the radiative corrections arises from missing higher-order terms, beyond NLO. The relative size of these corrections is assumed to be of order $(\alpha/\pi)\ln(Q^2/m_e^2)$, which we conservatively take as 10\%~\cite{Czyz:2025pc}. Furthermore, the radiative corrections have been shown to significantly depend on the specific event selection criteria~\cite{Czyz:2018jpp}. The relative difference in the reconstructed event yield from simulations with and without NLO radiative effects~\cite{Czyz:2025pc} is found to be 10\%.  Thus, an average systematic uncertainty of 10\% of 10\%, or 1\%, is assigned.

Systematic effects due to potential peaking background contributions are estimated by subtracting the background of $e^+e^-\to e^+e^-\pi^0\pi^0$, simulated with {\sc Ekhara 3.1}~\cite{Czyz:2018jpp,Danilkin:2018qfn}, from data before applying the standard fit procedure. Possible effects due to the subtraction of non-peaking background are tested by varying the fit range by $\pm10\,\text{MeV}$, and by changing the shape of the background contribution to a polynomial of third order. Deviations in the event yield from each variation from the nominal result range from $0.9\%$ to $35\%$ and are assigned as the systematic uncertainties. Like the $Q^2$ resolution effects discussed above, the largest values occur for the first and the last two $Q^2$ bins.

The contributions to the systematic uncertainty of the Born cross section are summed in quadrature. According to error propagation, the relative uncertainty of the TFF is half of the Born cross section uncertainty.

\begin{table}[htb]
 \caption{\label{tab:final} Number of background-subtracted signal events $N_{\rm sig}$, reconstruction efficiency $\varepsilon$, radiative correction factor $(1+\delta)$, and the point-like differential cross section $\frac{\textrm{d}\sigma_{\rm WZW}}{\textrm{d}Q^2}$ from MC simulation~\cite{Czyz:2018jpp}. Also shown are the extracted transition form factor (TFF) $|F(Q^2)|$ and its product with the central value of the momentum transfer $Q^2$. The first uncertainty is statistical, and the second is systematic.}
 \begin{center}
  \begin{small}
  \resizebox{1.0\columnwidth}{!}{
   \begin{tabular}{c||c|c|c|c||c|c}
$Q^2 [{\rm GeV}^2]$ & $N_{\rm sig}$ & $\varepsilon$ & $(1+\delta)$ & $\frac{\textrm{d}\sigma_{\rm WZW}}{\textrm{d}Q^2}$ [pb] & $|F(Q^2)| [{\rm GeV}^{-1}]$ & $Q^2\cdot |F(Q^2)| [{\rm GeV}]$ \\
\hline
$0.2 - 0.3$   & $  31 \pm 13$ & $0.0070 \pm 0.0005$ & $0.245 \pm 0.002$ & $2365   \pm 7  $ & $0.167 \pm 0.034 \pm 0.024$ & $0.042 \pm 0.009 \pm 0.006$ \\
$0.3 - 0.4$   & $ 508 \pm 36$ & $0.079  \pm 0.001 $ & $0.528 \pm 0.003$ & $1481   \pm 5  $ & $0.172 \pm 0.006 \pm 0.007$ & $0.060 \pm 0.002 \pm 0.002$ \\
$0.4 - 0.5$   & $1410 \pm 48$ & $0.205  \pm 0.001 $ & $0.939 \pm 0.005$ & $1039   \pm 4  $ & $0.159 \pm 0.003 \pm 0.004$ & $0.072 \pm 0.001 \pm 0.002$ \\
$0.5 - 0.6$   & $1358 \pm 44$ & $0.307  \pm 0.002 $ & $0.985 \pm 0.005$ & $ 775   \pm 3  $ & $0.144 \pm 0.002 \pm 0.003$ & $0.079 \pm 0.001 \pm 0.001$ \\
$0.6 - 0.7$   & $ 948 \pm 38$ & $0.318  \pm 0.002 $ & $0.979 \pm 0.006$ & $ 603   \pm 3  $ & $0.135 \pm 0.003 \pm 0.003$ & $0.088 \pm 0.002 \pm 0.002$ \\
$0.7 - 0.8$   & $ 630 \pm 30$ & $0.303  \pm 0.002 $ & $0.968 \pm 0.006$ & $ 480   \pm 2  $ & $0.127 \pm 0.003 \pm 0.004$ & $0.095 \pm 0.002 \pm 0.003$ \\
$0.8 - 0.9$   & $ 397 \pm 25$ & $0.278  \pm 0.002 $ & $0.979 \pm 0.007$ & $ 397   \pm 2  $ & $0.115 \pm 0.004 \pm 0.002$ & $0.098 \pm 0.003 \pm 0.002$ \\
$0.9 - 1.0$   & $ 298 \pm 22$ & $0.258  \pm 0.002 $ & $0.993 \pm 0.007$ & $ 332   \pm 2  $ & $0.112 \pm 0.004 \pm 0.003$ & $0.106 \pm 0.004 \pm 0.003$ \\
$1.0 - 1.15$  & $ 308 \pm 25$ & $0.250  \pm 0.002 $ & $0.992 \pm 0.006$ & $ 272   \pm 1  $ & $0.105 \pm 0.004 \pm 0.002$ & $0.112 \pm 0.005 \pm 0.003$ \\
$1.15 - 1.3$  & $ 276 \pm 21$ & $0.337  \pm 0.002 $ & $0.984 \pm 0.006$ & $ 218   \pm 1  $ & $0.096 \pm 0.004 \pm 0.003$ & $0.117 \pm 0.004 \pm 0.003$ \\
$1.3  - 1.45$ & $ 222 \pm 19$ & $0.372  \pm 0.002 $ & $0.999 \pm 0.007$ & $ 179   \pm 1  $ & $0.089 \pm 0.004 \pm 0.002$ & $0.123 \pm 0.005 \pm 0.003$ \\
$1.45 - 1.6$  & $ 156 \pm 16$ & $0.373  \pm 0.002 $ & $0.991 \pm 0.007$ & $ 147.5 \pm 0.8$ & $0.083 \pm 0.004 \pm 0.003$ & $0.126 \pm 0.006 \pm 0.005$ \\
$1.6 - 1.8$   & $ 141 \pm 16$ & $0.371  \pm 0.002 $ & $0.995 \pm 0.006$ & $ 123.4 \pm 0.6$ & $0.075 \pm 0.004 \pm 0.002$ & $0.127 \pm 0.007 \pm 0.003$ \\
$1.8 - 2.0$   & $  99 \pm 14$ & $0.369  \pm 0.002 $ & $0.997 \pm 0.007$ & $ 100.7 \pm 0.5$ & $0.069 \pm 0.005 \pm 0.004$ & $0.132 \pm 0.009 \pm 0.008$ \\
$2.0 - 2.2$   & $  72 \pm 12$ & $0.368  \pm 0.002 $ & $1.000 \pm 0.007$ & $  84.1 \pm 0.5$ & $0.065 \pm 0.005 \pm 0.005$ & $0.136 \pm 0.011 \pm 0.010$ \\
$2.2 - 2.4$   & $  59 \pm 14$ & $0.364  \pm 0.003 $ & $1.004 \pm 0.008$ & $  71.3 \pm 0.4$ & $0.064 \pm 0.007 \pm 0.003$ & $0.147 \pm 0.020 \pm 0.006$ \\
$2.4 - 2.6$   & $  28 \pm  9$ & $0.355  \pm 0.002 $ & $0.999 \pm 0.008$ & $  60.0 \pm 0.4$ & $0.049 \pm 0.008 \pm 0.003$ & $0.122 \pm 0.020 \pm 0.007$ \\
$2.6 - 2.8$   & $  38 \pm  9$ & $0.356  \pm 0.003 $ & $0.998 \pm 0.009$ & $  51.4 \pm 0.3$ & $0.061 \pm 0.007 \pm 0.004$ & $0.165 \pm 0.020 \pm 0.010$ \\
$2.8 - 3.1$   & $  13 \pm  8$ & $0.344  \pm 0.002 $ & $0.999 \pm 0.007$ & $  43.3 \pm 0.2$ & $0.032 \pm 0.010 \pm 0.006$ & $0.095 \pm 0.029 \pm 0.017$  \\
$3.1 - 3.5$   & $  12 \pm  8$ & $0.339  \pm 0.002 $ & $0.997 \pm 0.006$ & $  34.7 \pm 0.2$ & $0.030 \pm 0.010 \pm 0.008$ & $0.100 \pm 0.034 \pm 0.026$ \\
   \end{tabular}
   }
  \end{small}
 \end{center}
\end{table}

\section{Summary}

\begin{figure}[htb]
 \centerline{\includegraphics[width=0.9\textwidth]{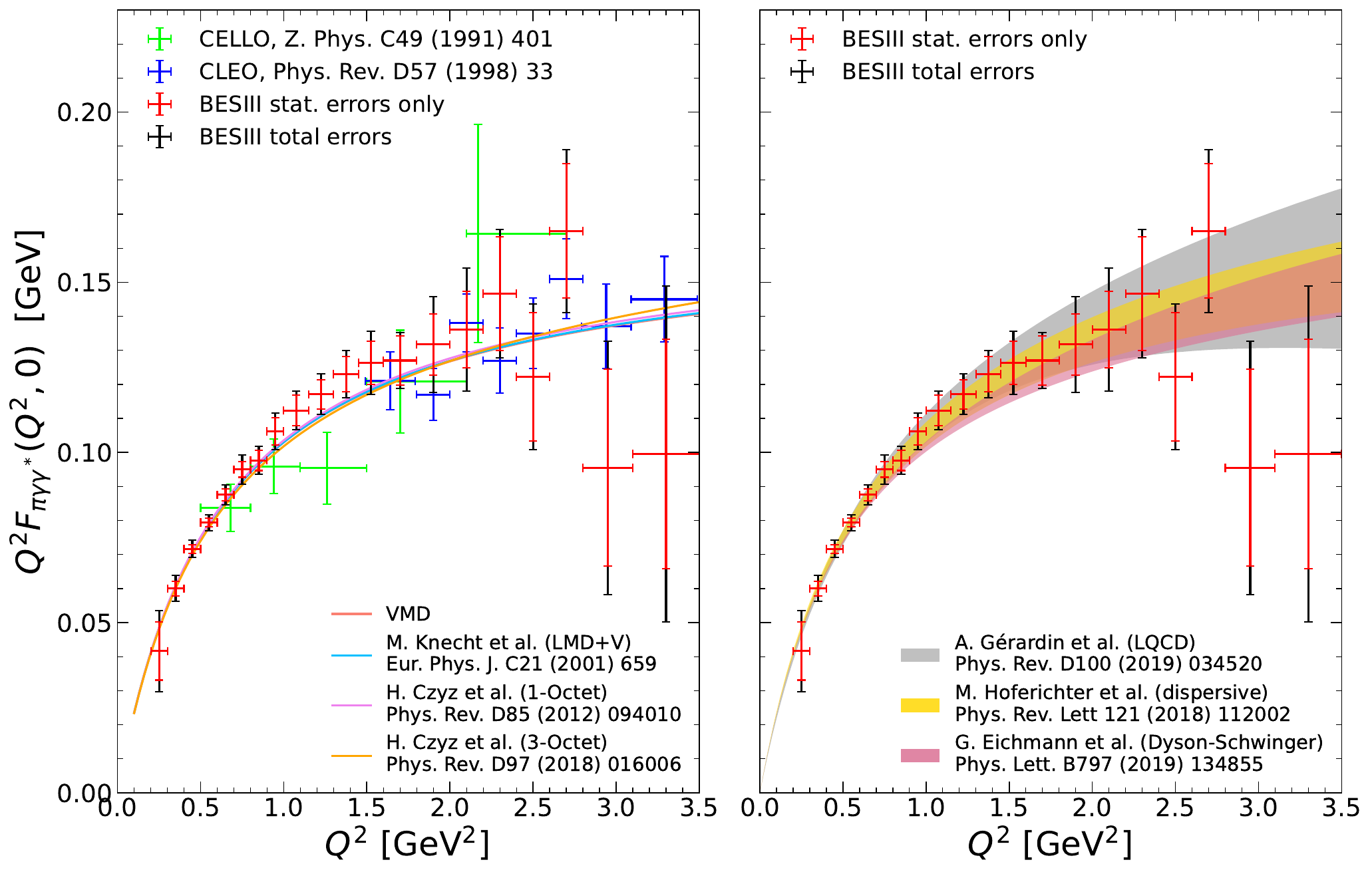}}
 \caption{\label{fig:final}Result of the $\pi^0$ TFF measurement with statistical (red error bars) and total uncertainties (black error bars) compared to previous measurements and hadronic models (left panel) and predictions from dispersive, lattice QCD, and Dyson-Schwinger theory calculations.}
\end{figure}

The $\pi^0$ TFF is determined from a measurement of the Born cross section of the two-photon fusion reaction $e^+e^- \to e^+e^- \pi^0$, using $2.93\,\text{fb}^{-1}$ of data taken at $\sqrt{s}=3.773\,\text{GeV}$ with the BESIII detector. A single-tag analysis is applied to investigate the $Q^2$ dependence of the TFF in the range $0.2 \leq Q^2 \leq 3.5\,\text{GeV}^2$. As illustrated in Fig.~\ref{fig:final}, the precision of this measurement is consistent with previous results at $Q^2\approx2\,\text{GeV}^2$ and surpasses earlier measurements in both precision and $Q^2$ coverage towards lower values. The result is in good agreement with various theoretical predictions. However, hadronic models~\cite{Knecht:2001xc,Czyz:2012nq,Czyz:2017veo} show a deviation of approximately $1\sigma$ in the range $0.9\leq Q^2\leq 1.5\,\text{GeV}^2$, consistent with a similar discrepancy observed in the CELLO~\cite{Behrend:1990sr} results in the same $Q^2$ range. Better agreement is observed with the calculations in dispersive~\cite{Hoferichter:2018dmo} (yellow band) and in lattice QCD~\cite{Gerardin:2019vio} (gray band) theory. The Dyson-Schwinger approach~\cite{Eichmann:2019tjk} (magenta band) yields slightly smaller values, which deviate by less than $1\sigma$ from the measurement. Since the latter three approaches provide vital input to the SM prediction of $a_\mu$~\cite{Aoyama:2020ynm}, their validation by the experimental result presented in this Letter is an important contribution.

The current accuracy is dominated by statistics. The BESIII collaboration finished data taking of a $20.3\,\text{fb}^{-1}$ data set at $\sqrt{s}=3.773\,\text{GeV}$ in 2024. Applying the current analysis to the full data set will significantly enhance the precision of the $\pi^0$ TFF determination. The increased statistics will also enable studies of the TFF as a function of both virtualities $|F(Q_1^2,Q_2^2)|$.

%% Saved at => 2025-02-13
\vspace*{2em}\noindent\textbf{Acknowledgment}

The BESIII Collaboration thanks the staff of BEPCII (https://cstr.cn/31109.02.BEPC) and the IHEP computing center for their strong support. This work is supported in part by National Key R\&D Program of China under Contracts Nos. 2023YFA1606000, 2023YFA1606704; National Natural Science Foundation of China (NSFC) under Contracts Nos. 11635010, 11935015, 11935016, 11935018, 12025502, 12035009, 12035013, 12061131003, 12192260, 12192261, 12192262, 12192263, 12192264, 12192265, 12221005, 12225509, 12235017, 12361141819; the Chinese Academy of Sciences (CAS) Large-Scale Scientific Facility Program; CAS under Contract No. YSBR-101; 100 Talents Program of CAS; The Institute of Nuclear and Particle Physics (INPAC) and Shanghai Key Laboratory for Particle Physics and Cosmology; German Research Foundation DFG under Contract No. FOR5327; Istituto Nazionale di Fisica Nucleare, Italy; Knut and Alice Wallenberg Foundation under Contracts Nos. 2021.0174, 2021.0299; Ministry of Development of Turkey under Contract No. DPT2006K-120470; National Research Foundation of Korea under Contract No. NRF-2022R1A2C1092335; National Science and Technology fund of Mongolia; National Science Research and Innovation Fund (NSRF) via the Program Management Unit for Human Resources \& Institutional Development, Research and Innovation of Thailand under Contract No. B50G670107; Polish National Science Centre under Contract No. 2024/53/B/ST2/00975; Swedish Research Council under Contract No. 2019.04595; U. S. Department of Energy under Contract No. DE-FG02-05ER41374

%        Example added sentence: This paper is also supported by the NSFC under Contract Nos. 10805053, 10979059, ....National Natural Science Foundation of China (NSFC), 10805053, PWANational Natural Science Foundation of China (NSFC), 10979059, Lund弦碎裂强子化模型及其通用强子产生器研究National Natural Science Foundation of China (NSFC), 10775075, National Natural Science Foundation of China (NSFC), 10979012, baryonsNational Natural Science Foundation of China (NSFC), 10979038, charmoniumNational Natural Science Foundation of China (NSFC), 10905034, psi(2S)->B BbarNational Natural Science Foundation of China (NSFC), 10975093, D 介子National Natural Science Foundation of China (NSFC), 10979033, psi(2S)->VPNational Natural Science Foundation of China (NSFC), 10979058, hcNational Natural Science Foundation of China (NSFC), 10975143, charmonium rare decays
%% ends here %%



\bibliography{literatur.bib}




%%\bibitem{}

%%\end{thebibliography}
\end{document}